\documentclass[doublecol,linenumbers]{epl2}
\usepackage[english]{babel}
\usepackage{blindtext}
\usepackage{amsmath}
\usepackage{array}
\usepackage{graphicx}
\usepackage{color}
\usepackage{makeidx}
\usepackage[latin1]{inputenc}
\usepackage{amssymb}
\usepackage{ngerman}
\usepackage{bm} 
\usepackage{subfigure} 
\usepackage{cite}

\newcommand{\D}{\textrm{d}}
\newcommand{\I}{\textrm{i}}

\newcommand{\expval}[1]{\langle #1 \rangle}
\newcommand{\nn}{\nonumber\\}

\newcommand{\f}[1]{\mbox{\boldmath$#1$}}

\newcommand{\abs}[1]{{\left| #1 \right|}}
\newcommand{\trace}[1]{{\rm Tr}\left\{ #1 \right\}}

\title{Steady-state thermodynamics of non-interacting transport beyond weak coupling}
\shorttitle{Thermodynamics of transport beyond weak coupling}

\date{\today}

\author{Gabriel E. Topp\inst{1}\thanks{E-mail: \email{fizztopp@gmail.com}} \and Tobias Brandes\inst{1} \and Gernot Schaller\inst{1}\thanks{E-mail: \email{gernot.schaller@tu-berlin.de}}}
\shortauthor{G. E. Topp, T. Brandes, and G. Schaller}
\institute{
\inst{1} Institut f"ur Theoretische Physik - Technische Universit"at Berlin, Hardenbergstrasse 36, D-10623 Berlin, Germany
}


\abstract{
We investigate the thermodynamics of simple (non-interacting) transport models beyond the scope of weak coupling. 
For a single fermionic or bosonic level -- tunnel-coupled to two reservoirs -- exact expressions for the stationary matter and energy current 
are derived from the solutions of the Heisenberg equations of motion.
The positivity of the steady-state entropy production rate is demonstrated explicitly.
Finally, for a configuration in which particles are pumped upwards in chemical potential by a 
downward temperature gradient, we demonstrate that the thermodynamic efficiency of this process
decreases when the coupling strength between system and reservoirs is increased, as a direct consequence of the 
loss of a tight coupling between energy and matter currents. 
}

\begin{document}

\maketitle

\section{Introduction}

Thermodynamic studies on nanoscale processes have been investigated intensively over the last years. 
For nanosystems, the traditional picture of a Carnot engine that is connected in an alternating fashion to different
thermal reservoirs is often replaced by stationary currents flowing through a quantum system continuously connecting 
such reservoirs~\cite{esposito2010a}.
When the coupling between system and environments is sufficiently weak, a master equation approach is well applicable. Within standard techniques~\cite{breuer2002} it is possible to derive thermodynamically consistent master equations of Lindblad form,
which lead to a positive entropy production~\cite{lindblad1975a} and thus obey the laws of thermodynamics.

With some exceptions~\cite{schaller2013a}, the thermodynamic study of such master equations is usually limited to the weak coupling regime between the system
and reservoirs~\cite{schaller2014}.
Within such master equation descriptions, the energy levels of the system become infinitely sharp.
In particular for a single intrinsic transition frequency, every particle that passes through the system must then carry
a defined amount of energy, resulting in a direct proportionality of the matter and the energy flow (termed {\em tight coupling}). 

More recently, the fate of thermodynamic bounds has also been investigated beyond master equation approaches~\cite{esposito2010b,tome2012a}.
However, many studies were limited to the case of a single reservoir that does not support stationary currents~\cite{morozov2012a}.

In this paper, we will consider stationary transport through either a bosonic or fermionic site between two reservoirs of similar type.
The weak-coupling results are of course faithfully reproduced but for the particular problems we will be mainly interested in the
strong-coupling regime.


\section{The simplest transport model}\label{SEC:transport_model}

To perform our thermodynamic studies we investigate the simplest conceivable models of quantum transport. 
To support a stationary current, such models should encompass (at least) two reservoirs, tunnel-coupled by a simple system
that locally has only a single transition frequency.
The Hamiltonian is thus of the form
\begin{eqnarray}\label{eq:hamiltonian}
\bm{H} &=& \epsilon \bm{d}^\dagger \bm{d}  + \sum_{k\alpha}\omega_{k \alpha} \bm{c}_{k \alpha}^\dagger \bm{c}_{k\alpha} \nn
	&\quad& + \sum_{k \alpha}\left[t_{k\alpha}\bm{d} \bm{c}_{k\alpha}^\dagger \pm t^*_{k \alpha}\bm{d}^\dagger \bm{c}_{k\alpha}\right]\,,
\end{eqnarray}
where $\epsilon$ denotes the system-intrinsic transition frequency, $\omega_{k\alpha}$ the frequency of mode $k$ of reservoir $\alpha\in\{\text{L},\text{R}\}$, and
$t_{k\alpha}$ the corresponding tunneling amplitudes between system and reservoir.
The annihilation operators for the system ($\bm{d}$) and the reservoirs ($\bm{c}_{k\alpha}$) can be of either bosonic (upper sign) or fermionic (lower sign) nature, 
manifested in corresponding commutation or anticommutation relations, respectively.
In the following expressions, we will adopt this convention, i.e., in case of differences between bosons and fermions the upper sign will hold for the bosonic transport model, 
and the lower for the fermionic counterpart.

We will solve for the resulting global dynamics, using only that initially the reservoirs are prepared in
grand-canonical equilibrium states, such that 
the initial state of the compound system is given by
\begin{eqnarray}\label{eq:initstate}
	\rho_0 = \rho_\text{S}^0 \bigotimes\limits_{\alpha\in\{\text{L},\text{R}\}} \frac{e^{-\beta_\alpha(H_\alpha - \mu_\alpha N_\alpha)}}{\trace{e^{-\beta_\alpha(H_\alpha - \mu_\alpha N_\alpha)}}}\,,
\end{eqnarray}
with $\trace{\rho_\text{S}^0}=1$.
Here, $\beta_\alpha$ and $\mu_\alpha$ denote the initial inverse temperature and chemical potential of reservoir $\alpha$, 
$H_\alpha$ its Hamiltonian -- cf. Eq.~(\ref{eq:hamiltonian}) -- and 
$N_\alpha = \sum_k \bm{c}_{k\alpha}^\dagger \bm{c}_{k\alpha}$ the corresponding particle number operator.
Obviously, there is no initial entanglement between the system and the environment.
Our treatment does not require the amplitudes $t_{k\alpha}$ to be small.
However, we remark that in general, for large amplitudes the distinction between 
system and reservoir becomes somewhat fuzzy, since some eigenstates of the Hamiltonian will extend over
both system and reservoir.
For example, it is in principle possible that energy contained in the interaction Hamiltonians then 
contributes non-negligibly to steady-state transport.
We will see later that for our setup this is not the case, enforced by the initial condition~(\ref{eq:initstate})
and the assumption of a continuous spectral coupling density.

\section{Equation-of-motion method}\label{SEC:equation_of_motion}

In the equation-of-motion-method, one simply solves the Heisenberg picture dynamics
$\frac{\D}{\D t}\bm{A} = \I \left[\bm{H},\bm{A}\right]$ for the bosonic or fermionic operators
generated by Eq.~(\ref{eq:hamiltonian}) exactly.
Afterwards, physical observables of interest can be obtained from the time-evolved operators.

For a quadratic Hamiltonian without interactions as considered here, the Heisenberg equations of motion just yield a linear set of first order differential equations~\cite{yang2013a}
\begin{eqnarray}\label{eqs:operators}
\dot{\bm{d}} &=& -\I \epsilon \bm{d} \mp \I \sum_{k\alpha} t^*_{k\alpha} \bm{c}_{k\alpha}\,,\nn
\dot{\bm{c}}_{k\alpha} &=& -\I \omega_{k\alpha} \bm{c}_{k\alpha} \mp \I t_{k\alpha}\bm{d}\,,
\end{eqnarray} 
and similarly for the creation operators, where the bosonic ($-$) or fermionic ($+$) nature of the operators just manifests in the sign.
We note that in case of interactions present, the set will not close and approximations must be applied for further treatment.
Similar to coupled equations for the evolution of amplitudes~\cite{gurvitz1991a}, this closed set of infinitely many 
differential equations can be transformed to an algebraic set of equations~\cite{schaller2009a} using a Laplace transform
${\cal L}\left\{\bm{d}\right\} \equiv \hat{\bm{d}}(z) = \int_0^\infty e^{-zt}\bm{d}(t)\D t$, which can then be solved explicitly in terms of the initial operators
(denoted by $d$ and $c_{k\alpha}$, respectively).
For later convenience, we state the result for 
$\hat{\bm{d}}(z)$ and $\hat{\bm{c}}_{k\alpha}^\dagger(z)$ explicitly
\begin{eqnarray}\label{eq:oplapla}
\hat{\bm{d}}(z)	&=& \frac{d}{z+\I \epsilon+\sum\limits_{k\alpha} \frac{\abs{t_{k\alpha}}^2}{z+\I\omega_{k\alpha}}}\nn
&&\mp\I \sum_{k\alpha} \frac{t_{k\alpha}^*  c_{k\alpha}}{(z+\I\omega_{k\alpha})\left(z+\I\epsilon+\sum\limits_{k'\alpha'} \frac{\abs{t_{k'\alpha'}}^2}{z+\I\omega_{k'\alpha'}}\right)}\,,\nn
\hat{\bm{c}}_{k\alpha}^\dagger(z) &=& \frac{c_{k\alpha}^\dagger}{z-\I\omega_{k\alpha}}
\pm\frac{\I t_{k\alpha}^* d^\dagger}{\left(z-\I\omega_{k\alpha}\right)\left(z-\I\epsilon+\sum\limits_{k'\alpha'} \frac{\abs{t_{k'\alpha'}}^2}{z-\I\omega_{k'\alpha'}}\right)}\nn
&&+ \sum_{k'\alpha'} \frac{t_{k\alpha}^* t_{k'\alpha'} }{\left(z-\I\omega_{k\alpha}\right)\left(z-\I\omega_{k'\alpha'}\right)}\times\nn
&&\qquad\times \frac{c_{k'\alpha'}^\dagger}{\left(z-\I\epsilon+\sum\limits_{k''\alpha''} \frac{\abs{t_{k''\alpha''}}^2}{z-\I\omega_{k''\alpha''}}\right)}\,.
\end{eqnarray}
Now, exact solutions can in principle be obtained by inverting the Laplace transform, which in the general case is typically tedious.
Since we want to consider steady state configurations, we must assume an infinite and continuous distribution of reservoir modes $k$ 
(finite-sized quantum systems evolve periodically).
Technically, this corresponds to the replacement of
\begin{eqnarray}
\sum_k \frac{\abs{t_{k\alpha}}^2}{z+\I\omega_{k\alpha}} &\to& \frac{1}{2\pi}\int\limits_0^\infty \frac{\Gamma_\alpha(\omega)}{z+\I\omega} \text{d}\omega \equiv C_\alpha(z)\,,\nn
\sum_k \frac{\abs{t_{k\alpha}}^2}{z-\I\omega_{k\alpha}} &\to& \frac{1}{2\pi}\int\limits_0^\infty \frac{\Gamma_\alpha(\omega)}{z-\I\omega} \text{d}\omega \equiv \bar{C}_\alpha(z)\,,
\end{eqnarray}
with the energy-dependent tunneling rate (or spectral coupling density) $\Gamma_\alpha(\omega) = 2\pi \sum_k \abs{t_{k\alpha}}^2 \delta(\omega-\omega_{k\alpha})$.
Finally, we mention that to obtain the solution for the hermitian conjugate operators, the Laplace transform variable should be kept invariant 
to remain consistent with our further calculations (meaning the same Laplace transform is applied to creation and annihilation operators in the time domain).

\section{The currents}\label{SEC:currents}
 
\begin{figure}[ht]
\begin{center}
\includegraphics[width=0.35\textwidth]{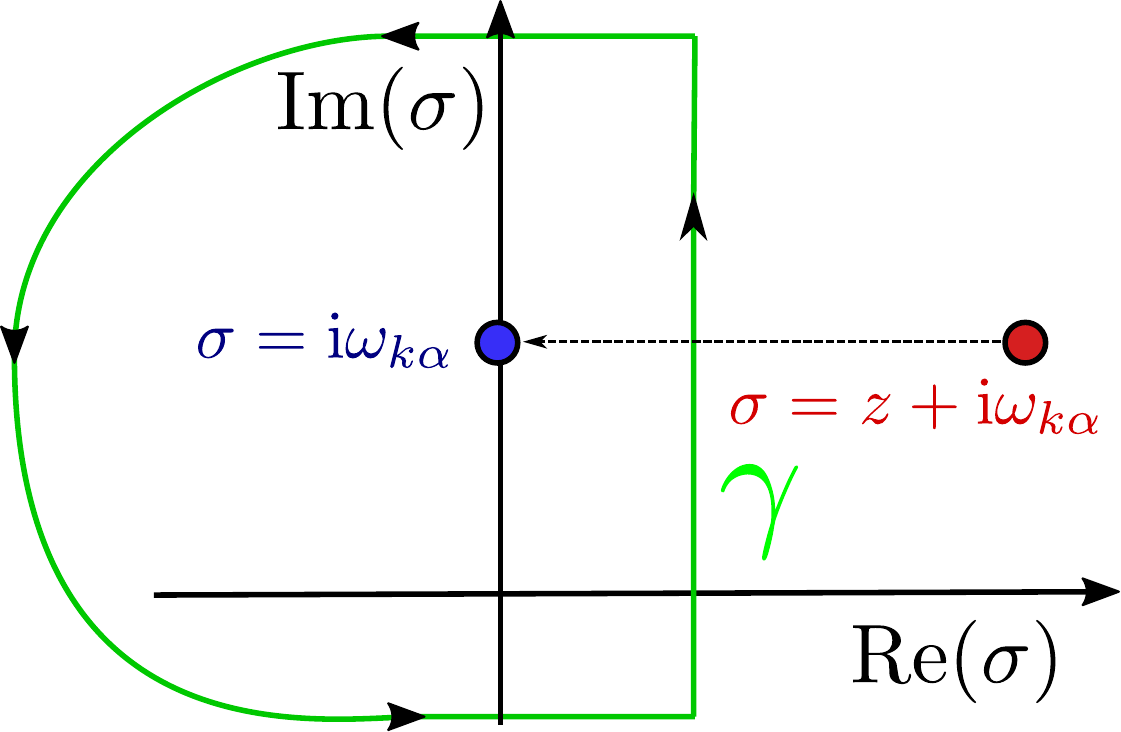}
\end{center}
\caption{\label{FIG:pole_sketch}(Color Online)
Sketch of the relevant poles arising in the expectation value $\expval{\hat{\bm{c}}_{k\text{R}}^\dagger(\sigma)\hat{\bm{d}}(z-\sigma)}$.
The contour parameter $\gamma$ must be chosen such that the poles of $\hat{\bm{c}}_{k\text{R}}^\dagger(\sigma)$ (blue) lie left and the poles
of $\hat{\bm{d}}(z-\sigma)$ (red) lie right to it.
When the limit $z\to 0$ is performed afterwards (thin dashed arrows), it becomes visible that in order to yield a long-term contribution, poles inside the integration contour
described by $\sigma-\I \omega=0$ can only contribute if they have a conjugate pole ($z-\sigma+\I \omega=0$) outside the contour.
In this case, when evaluating the residue one obtains
$\lim\limits_{z\to 0} \left.\frac{z}{z-\sigma+\I \omega}\right|_{\sigma=\I \omega} = 1$.
}
\end{figure}

We are interested in the long-term matter and energy currents entering the right lead and therefore compute the 
Heisenberg equation of motions for the particle number operator and the Hamiltonian of the right reservoir, 
yielding
\begin{eqnarray}\label{eq:dnde}
\dot{\bm{N}}_\text{R} &=& \mp\I \sum_k \left[ t_{k \text{R}}\bm{c}_{k \text{R}}^\dagger \bm{d} - \text{h.c.}\right]\,, \nn
\dot{\bm{H}}_\text{R} &=& \mp\I \sum_k \left[ t_{k \text{R}}\omega_{k \text{R}}\bm{c}_{k \text{R}}^\dagger \bm{d} - \text{h.c.}\right]\,. 
\end{eqnarray}
Computing expectation values and considering the long-time limit, 
we find that the steady-state matter and energy currents into the right reservoir can be calculated by convolution integrals in Laplace space
\begin{eqnarray}\label{eq:ImEmcon}
I_\text{M} &=& \lim_{t\to\infty} \expval{\dot{\bm{N}}_\text{R}} = \lim_{z\to 0} z \mathcal{L}\left\{\expval{\dot{\bm{N}}_{\text{R}}}\right\}\nn
&=& \mp\lim_{z\to 0} \frac{z}{2 \pi}\int\limits_{\gamma-\I\infty}^{\gamma+\I\infty}\sum_k
t_{k\text{R}}\expval{\hat{\bm{c}}_{k\text{R}}^\dagger(\sigma)\hat{\bm{d}}(z-\sigma)}\D\sigma + \text{h.c.}\,,\nn
I_\text{E} &=& \lim_{t\to\infty} \expval{\dot{\bm{H}}_\text{R}} = \lim_{z\to 0 } z \mathcal{L}\left\{\expval{\dot{\bm{H}}_{\text{R}}}\right\}\nn
&=& \mp\lim_{z\to 0} \frac{z}{2 \pi}\int\limits_{\gamma-\I\infty}^{\gamma+\I\infty}\sum_k\omega_{k\text{R}}
t_{k\text{R}}\left\expval{\hat{\bm{c}}_{k\text{R}}^\dagger(\sigma)\hat{\bm{d}}(z-\sigma)\right}\D\sigma\nn
&&+\text{h.c.}
\end{eqnarray} 
Here, the real constant $\gamma$ has to be chosen such that the poles of $\hat{\bm{c}}_{k\text{R}}^\dagger(\sigma)$ lie left to it and the 
remaining poles of ${\hat{\bm{d}}(z-\sigma)}$ are on the right of the integration contour as depicted in Fig.~\ref{FIG:pole_sketch}.

To evaluate the currents explicitly, we note that due to our condition of initial thermal equilibrium~(\ref{eq:initstate}), only few
terms may potentially contribute, since 
${\expval{c^{\dagger}_{k'\alpha'} c_{k\alpha}}= \delta_{kk'}\delta_{\alpha\alpha'}n_\alpha(\omega_k)}$  
with the Bose ($-$) and Fermi ($+$) distributions
\begin{eqnarray}\label{eq:fermibose}
n_\alpha(\omega_k) =  \frac{1}{e^{\beta_\alpha(\omega_{k\alpha}-\mu_\alpha)} \mp 1}\,.
\end{eqnarray}
We can evaluate these contributions with the residue theorem.

For simplicity, we first discuss the term arising from the combination $\expval{d^\dagger d}=n_0$.
Here, the residue arising from the pole at $\sigma=+\I\omega_{k\alpha}$ is multiplied by $z$ and thus vanishes as $z\to 0$.
Furthermore, the equation $\tilde\sigma-\I \epsilon + \bar{C}_\text{L}(\tilde\sigma)+\bar{C}_\text{R}(\tilde\sigma)=0$ can only be solved by a purely imaginary $\tilde\sigma=\I\sigma_y$
with (for $\epsilon>0$) $\sigma_y>0$.
Since we have that \mbox{$\lim\limits_{z\to 0}\left[z-\sigma+\I \epsilon + C_\text{L}(z-\sigma)+C_\text{R}(z-\sigma)\right]_{\sigma=\tilde\sigma}=0$}, we can invoke the 
rule of L'Hospital to find that
$\lim\limits_{z\to 0} \left.\frac{z}{z-\sigma+\I \epsilon + C_\text{L}(z-\sigma)+C_\text{R}(z-\sigma)}\right|_{\sigma=\tilde\sigma} 
= \frac{1}{1-\frac{1}{2\pi} \int_0^\infty \frac{\Gamma_\text{L}(\omega)+\Gamma_\text{R}(\omega)}{(\sigma_y - \omega)^2} \D\omega}=0$, 
where we have used that $\sigma_y>0$ leads to a divergence of the integral.
Therefore, as expected, the steady-state currents will not depend on the initial occupation of the system.

The same arguments can be applied to evaluate the other contributions. 
In particular, combining the first term in $\hat{\bm{c}}_{k\text{R}}^\dagger(\sigma)$ with the second term in $\hat{\bm{d}}(z-\sigma)$ we obtain
a stationary contribution to the matter current
\begin{eqnarray}\label{EQ:intm1}
I_\text{M}^{(1)} &=& -\int\limits_0^\infty \frac{\text{d}\omega}{2\pi} \frac{\Gamma_\text{R}(\omega) n_\text{R}(\omega)}{-\I \omega + \I \epsilon + \sum_\alpha \left[\frac{\Gamma_\alpha(\omega)}{2} + \I \Sigma_\alpha(\omega)\right]}\,,
\end{eqnarray}
for which we used
\begin{eqnarray}
\lim_{z \to 0^+} C_\alpha(z- \I \omega) = C_\alpha (-\I \omega) = \frac{\Gamma_\alpha(\omega)}{2} + \I \Sigma_\alpha(\omega)\,,
\end{eqnarray}
where $\Sigma_\alpha(\omega) = \frac{1}{2\pi}\mathcal{P} \int_{0}^{\infty}\frac{\Gamma_\alpha(\omega')}{\omega-\omega'}\D \omega'$ 
(with Cauchy principal value $\mathcal{P}$) represents a coupling-induced level renormalization (often called {\em Lamb shift}).
Combining the last term in $\hat{\bm{c}}_{k\text{R}}^\dagger(\sigma)$ with the second term in $\hat{\bm{d}}(z-\sigma)$ one obtains 
a second contribution
\begin{eqnarray}\label{EQ:intm2}
I_\text{M}^{(2)} &=&\int\limits_0^\infty \frac{\text{d}\omega}{2\pi} 
\frac{\bar{C}_\text{R}(\I\omega) \left[\Gamma_\text{L}(\omega) n_\text{L}(\omega) + \Gamma_\text{R}(\omega) n_\text{R}(\omega)\right]}
{\left(\frac{\Gamma(\omega)}{2}\right)^2 + \left[\omega-\epsilon-\Sigma(\omega)\right]^2}\,,\qquad
\end{eqnarray}
where $\Gamma(\omega) = \sum_\alpha \Gamma_\alpha(\omega)$ and in a similar fashion for $\Sigma(\omega)$.
To further separate real and imaginary parts we can use
$\lim\limits_{\sigma\to\I\omega} \bar{C}_\text{R}(\sigma) =\bar{C}_\text{R}(\I \omega) = \frac{\Gamma_\text{R}(\omega)}{2} -\I \Sigma_\text{R}(\omega)$.

Combining these expressions in the total matter current $I_\text{M} = 2 \Re(I_\text{M}^{(1)} + I_\text{M}^{(2)})$
and performing similar calculations for the energy current, we finally obtain for the steady-state currents the expressions
\begin{eqnarray}\label{eq:mcfecf}
I_\text{M} &=& \int\limits_{0}^{\infty} G_{\text{C}}(\omega)\left[n_\text{L}(\omega)-n_\text{R}(\omega)\right]	S_{\text{C}}(\omega) \D\omega \,, \nn
I_\text{E} &=& \int\limits_{0}^{\infty} \omega \cdot G_{\text{C}}(\omega)\left[n_\text{L}(\omega)-n_\text{R}(\omega)\right]S_{\text{C}}(\omega)\D\omega\,, 
\end{eqnarray}
with the factors
\begin{eqnarray}\label{eq:mcfecffactors}
G_{\text{C}}(\omega) &=& \frac{\Gamma_\text{L}(\omega)\Gamma_\text{R}(\omega)}{\Gamma(\omega)}\,, \nn	
S_{\text{C}}(\omega) &=& \frac{1}{\pi} \frac{\Gamma(\omega)/2 }{(\Gamma(\omega)/2)^2+\left[ \omega - \epsilon- \Sigma(\omega)\right]^2}\,,
\end{eqnarray}
and where $\Gamma(\omega) = \Gamma_\text{L}(\omega)+\Gamma_\text{R}(\omega)$ and $\Sigma(\omega) = \Sigma_\text{L}(\omega)+\Sigma_\text{R}(\omega)$.
The difference between bosonic and fermionic transport enters in the different distributions~(\ref{eq:fermibose}) and the fact that for bosons, the chemical
potentials must be negative (otherwise the total particle number does not converge), whereas they are unbounded for fermions.
Furthermore, also for fermions we assumed the reservoir frequencies $\omega_{k\alpha}$ to be positive. 
Relaxing this assumption would simply extend the lower bound in all integrals to $-\infty$.

The Landauer form of the current is well-known for purely fermionic~\cite{blanter2000a,haug2008,jin2010a,eich2014a}
and purely bosonic transport~\cite{ozpineci2001a,segal2003a}.
Structurally similar results hold for bosonic transport through a central spin~\cite{saito2013a,yang2014a}.

A nice feature of Eqns.~(\ref{eq:mcfecf}) and (\ref{eq:mcfecffactors}) is the common representation for both bosons and fermions.
The key advantage of this \textit{Landauer}-like representation \cite{Landauer1957,Meir1992,Meir1993} is that it is completely independent 
of the actual shape of $\Gamma_\alpha(\omega)$. 

For consistency, we note that in the weak-coupling limit $\Gamma_\alpha(\omega) \to 0$ (also implying $\Sigma(\omega)\to 0$), one 
of the factors becomes a Dirac-Delta function $S_{\text{C}}(\omega) \to {\delta(\omega - \epsilon - \Sigma(\omega)) \to \delta(\omega -\epsilon)}$, such
that the integrals collapse and the master equation results are reproduced
\begin{eqnarray}\label{eq:weak_coupling_current}
I_\text{M} &\to& \frac{\Gamma_\text{L}(\epsilon)\Gamma_\text{R}(\epsilon)}{\Gamma_\text{L}(\epsilon)+\Gamma_\text{R}(\epsilon)} \left[n_\text{L}(\epsilon)-n_\text{R}(\epsilon)\right]\,,\; I_\text{E} \to \epsilon I_\text{M}\,.\qquad
\end{eqnarray}

Assuming both strong coupling and flat tunneling rates $\Gamma_\alpha(\omega) \to \Gamma_\alpha$ [such that $S_{\rm C}(\omega) \to 2/(\pi \Gamma)$], 
we can obtain analytic results for the bosonic (upper sign) and fermionic (lower sign) matter and energy currents
\begin{eqnarray}\label{eq:strong_coupling_current}
I_\text{M} &\to& \alpha \left[\mp \frac{\ln\left(1\mp e^{\beta_\text{L} \mu_\text{L}}\right)}{\beta_\text{L}} \pm \frac{\ln\left(1\mp e^{\beta_\text{R} \mu_\text{R}}\right)}{\beta_\text{R}}\right]\,,\nn
I_\text{E} &\to& \alpha \left[\pm \frac{{\rm Li}_2(\pm e^{\beta_\text{L} \mu_\text{L}})}{\beta_\text{L}^2} \mp \frac{{\rm Li}_2(\pm e^{\beta_\text{R} \mu_\text{R}})}{\beta_\text{R}^2}\right]\,.
\end{eqnarray}
where $\alpha\equiv\frac{2}{\pi} \frac{\Gamma_\text{L} \Gamma_\text{R}}{\left(\Gamma_\text{L}+\Gamma_\text{R}\right)^2}$ and ${\rm Li}_2(x)$ denotes the polylog function.
Even simpler expressions arise for fermions when the integral is extended to the complete real axis (not shown).

For finite coupling strengths, the factor ${S_{\text{C}}(\omega)}$ encodes the coupling-induced level renormalization via ${\Sigma(\omega)}$. 
The integral over all energies $\omega$ in the expressions for the currents~(\ref{eq:mcfecf}) can be interpreted as a broadening of the 
system energy level, since all modes contribute to the transport, weighted by the tunneling rates. 
This broadening violates the tight-coupling of matter and energy fluxes.

We remark that the long-term energy content 
of the interaction Hamiltonian between system and the right lead can be extracted 
from the imaginary parts of the integrals in Eq.~(\ref{EQ:intm1}) and Eq.~(\ref{EQ:intm2}) 
via $\abs{\expval{\f{H}_{\I,R}(t)}} \to 2 \Im(I_\text{M}^{(1)} + I_\text{M}^{(2)})$.
Since these remain finite for large times, we conclude that the interaction energy does not 
contribute to the steady-state energy currents.
For finite-time statements however~\cite{wang2012c,wu2014a,esposito2015a,campisi2015a}, these contributions will matter.

\section{Thermodynamics}\label{SEC:thermodynamics}

\subsection{Steady-state entropy production}

When a system is coupled to different equilibrium environments, its change of entropy can be split in terms that have an 
interpretation as entropy flow and entropy production~\cite{esposito2010b}.
In particular, when a finite system reaches a steady-state (and hence its change of entropy vanishes), the internal
entropy production rate $\Delta_\text{i}\dot{S}$ must be balanced by the entropy flow $\Delta_\text{e} \dot{S}$ entering the system, 
\begin{eqnarray}
	\Delta_\text{i}\dot{S} 
	= -\Delta_\text{e} \dot{S} 
	=	-\sum_\alpha \beta_\alpha \dot{Q}_\alpha \ge  0.  \label{eq:ep}
\end{eqnarray}
Here, $\dot{Q}_\alpha$ denotes the heat current entering the system from reservoir $\alpha$. 
We note that positivity of {\em integrated} entropy production -- defined as difference between change of internal entropy and integrated entropy 
flow -- has been proven generally~\cite{esposito2010b,reeb2014a} and also at the level of individual trajectories~\cite{campisi2014a}.
Here, we will demonstrate explicitly that our transport scenario -- without changes in the basic thermodynamic definitions~\cite{esposito2015a} -- 
at steady state supports a positive entropy production {\em rate}~(\ref{eq:ep}). 
The proof is simple and general for Landauer-B\"uttiker transport~\cite{nenciu2007a}.
Using matter and energy conservation (one can of course explicitly check that at steady state the first law is obeyed), the heat currents entering the system are defined as
\begin{eqnarray}
\dot{Q}_\text{L} = +I_\text{E} -\mu_\text{L}I_\text{M}\,,\qquad
\dot{Q}_\text{R} = -I_\text{E} +\mu_\text{R}I_\text{M}\,.
\end{eqnarray} 
Altogether, the internal entropy production rate is thus of the form
\begin{eqnarray}\label{eq:entropy_production}
\Delta_\text{i}\dot{S}	
	= (\beta_\text{R}-\beta_\text{L})I_\text{E} + (\mu_\text{L}\beta_\text{L}-\mu_\text{R}\beta_\text{R})I_\text{M}\,.
\end{eqnarray}

Inserting the integral expressions for the currents to the equation above yields
\begin{eqnarray}\label{eq:intent}
\Delta_\text{i}\dot{S}&=&
\int \limits_{0}^{\infty}\chi(\omega) G_{\text{C}}(\omega)S_{\text{C}}(\omega)\D\omega\,,\\
\chi(\omega) &=& \left[\omega(\beta_\text{R}-\beta_\text{L})+ (\mu_\text{L}\beta_\text{L}-\mu_\text{R}\beta_\text{R})\right] \left[n_\text{L}(\omega)-n_\text{R}(\omega)\right]\,.\nonumber
\end{eqnarray}
As one obviously has ${G_{\text{C}}(\omega)S_{\text{C}}(\omega) \ge 0 \quad \forall \omega}$, since ${\Gamma(\omega) \ge 0}$, it is sufficient to show that
$\chi(\omega) \ge  0$.
In the trivial case of equal temperatures and chemical potentials ($n_\text{L}(\omega) = n_\text{R}(\omega)$), 
all currents vanish and the entropy production is zero. 
Only for different temperatures, a non-trivial root of $\chi(\omega)$ at $\omega_0 = \frac{\beta_\text{L}\mu_\text{L}-\beta_\text{R}\mu_\text{R}}{\beta_\text{L}-\beta_\text{R}}$ exists with
\begin{eqnarray}
\left.\frac{\D}{\D \omega}\chi(\omega)\right|_{\omega = \omega_0} &=& 0\,, \\
\left.\frac{\D^2}{\D \omega^2}\chi(\omega)\right|_{\omega = \omega_0} 
	&=& \frac{(\beta_\text{L}-\beta_\text{R})^2}{\cosh\left[\frac{\beta_\text{L}\beta_\text{R}}{\beta_\text{L}-\beta_\text{R}}(\mu_\text{L}-\mu_\text{R})\right]\mp 1} \ge 0\,.\nonumber
\end{eqnarray} 
Consequently, the root at $\omega_0$ also corresponds to a global minimum. 
As $\chi^\pm(\omega)$ is continuously differentiable for finite temperatures and vanishes in the 
limit of $\omega\to\infty$, we obtain that $\chi(\omega) \ge 0$.
Non-negativity of the integrand in Eq.~(\ref{eq:intent}) along the real axis thus also implies a positive 
internal entropy production rate at steady state.
A similar proof would hold if the Bose-Einstein or Fermi-Dirac distributions were replaced by Boltzmann factors.

We mention that this second law automatically implies that at equal temperatures matter will always flow from reservoirs with large to the one with small chemical potential
and that at equal chemical potentials heat will always flow from hot to cold reservoirs.

\subsection{Nanothermal engine}

Now we consider a configuration in which our transport setup may function as a simple nanothermal engine. 
For fermions, exactly this kind of model has been considered before in the weak-coupling regime~\cite{M.Esposito2009}. 
Assuming that the temperature of the left lead is smaller than that of the right $\beta_\text{L}>\beta_\text{R}$, whereas the
potential gradient $V = \mu_\text{L}-\mu_\text{R} > 0$ is tilted in the other direction, it is in certain parameter regions 
possible to induce a current against the potential gradient ($I_\text{M}<0$), such that the generated power
\begin{eqnarray}
P= -I_\text{M}(\mu_\text{L}-\mu_\text{R})
\end{eqnarray} 
becomes positive.
For positive power, the thermodynamic efficiency of this process is defined as the ratio between generated power 
and the heat current entering the system from the hot (right) reservoir,
\begin{eqnarray}\label{eq:eff}
\eta &=& \frac{P}{\dot{Q}_\text{R}}\Theta(P) \nn 
&=& \frac{I_\text{M}(\mu_\text{L}-\mu_\text{R})}{I_\text{E} -\mu_\text{R} I_\text{M}}\Theta(-I_\text{M}(\mu_\text{L}-\mu_\text{R}))\,,
\end{eqnarray}
where we have introduced the Heaviside-function to enforce the correct regime
(finite-time realizations~\cite{schaller2014b,delcampo2014a,campisi2015a} would have to relate work output and heat input).
From positivity of Eq.~(\ref{eq:entropy_production}) we can deduce that the efficiency is always 
upper-bounded by Carnot efficiency $\eta \le \eta_\text{C} = 1 - T_\text{L}/T_\text{R}$.

\subsection{Loss of tight-coupling}

In the weak-coupling limit -- compare Eq.~(\ref{eq:weak_coupling_current}) -- 
there is only a single transition frequency $\epsilon$ in the system, and
the currents become tightly coupled $I_\text{E} = \epsilon I_\text{M}$. 
Then, one can explicitly show that Carnot efficiency is actually reached when the current (and hence, also the power) vanishes.
Due to the fact that the power at maximum efficiency is zero, it has become customary to consider the efficiency at maximum power output instead.
Since this still goes beyond the linear response regime, this becomes an optimization problem 
that can in general not be solved analytically but requires numerical approaches.

In contrast, in the strong-coupling regime -- compare Eq.~(\ref{eq:strong_coupling_current}) -- 
one can numerically verify that the power at maximum efficiency does not need to be significantly smaller
than the maximum power.

As we will show in the following, beyond the weak-coupling regime, the power generated at maximum efficiency remains finite.
%
With the aim of a symmetric and consistent description for both bosons and fermions, we introduce the new variables
\begin{eqnarray}\label{eq:new_variables}
2 \beta^{-1} &=& \beta_\text{L}^{-1} + \beta_\text{R}^{-1}\,, \qquad \Delta\beta = \beta_\text{L} - \beta_\text{R}\nn
2 \mu^{-1} &=& \mu_\text{L}^{-1} + \mu_\text{R}^{-1}\,, \qquad \Delta\mu = \mu_\text{L} - \mu_\text{R}\,.
\end{eqnarray}
Thus, with the correct choice of the parameters $\beta$ and $\mu$, we are able to vary the (inverse) temperature gradient $\Delta\beta$ 
and the voltage bias $\Delta\mu$ over the complete real axis. 
In particular for the bosonic model, this representation is useful to ensurse negative chemical potentials throughout.
\begin{figure*}[ht]
\begin{tabular}{ccc}
\includegraphics[width=0.31\textwidth]{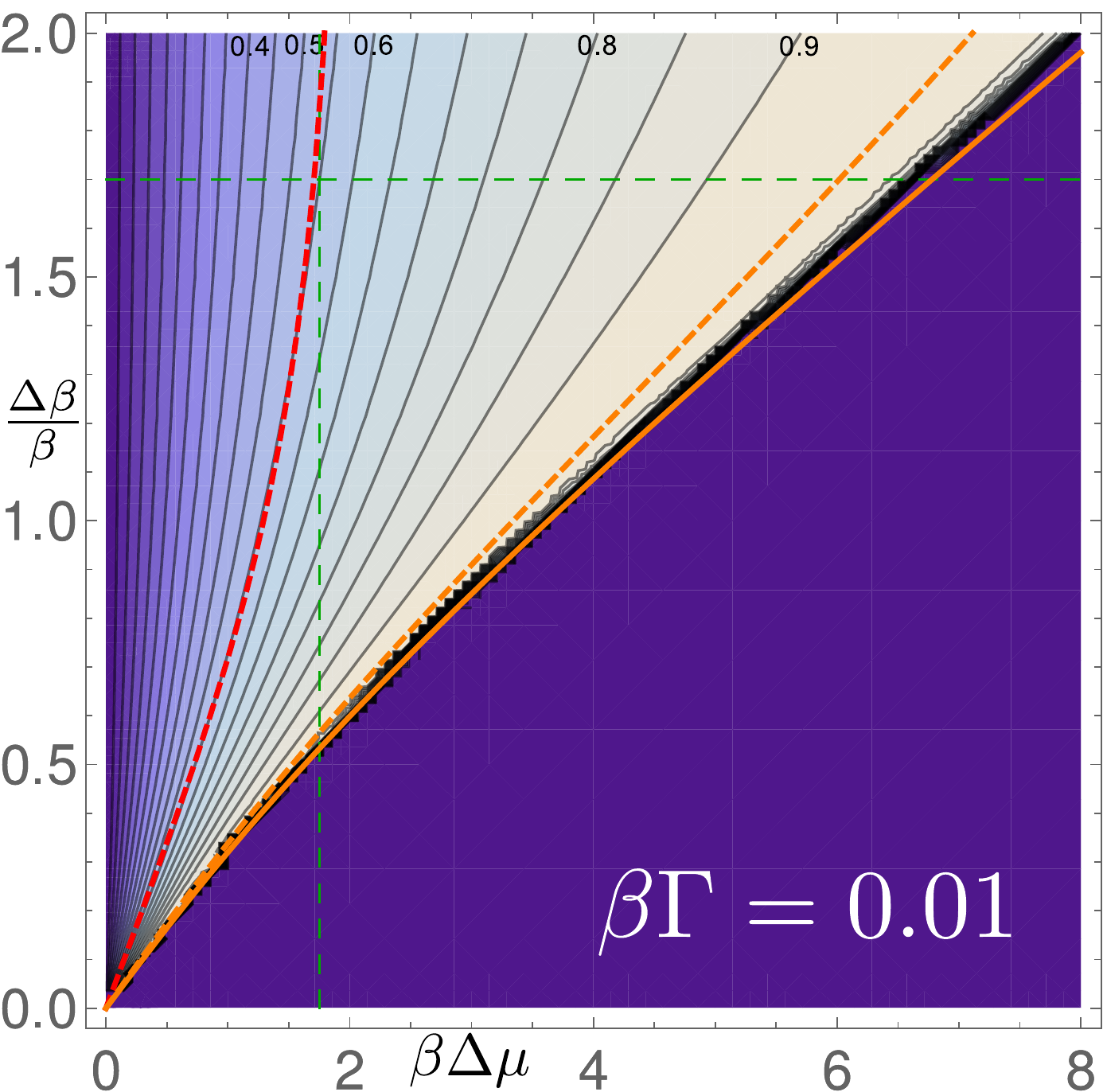} &
\includegraphics[width=0.31\textwidth]{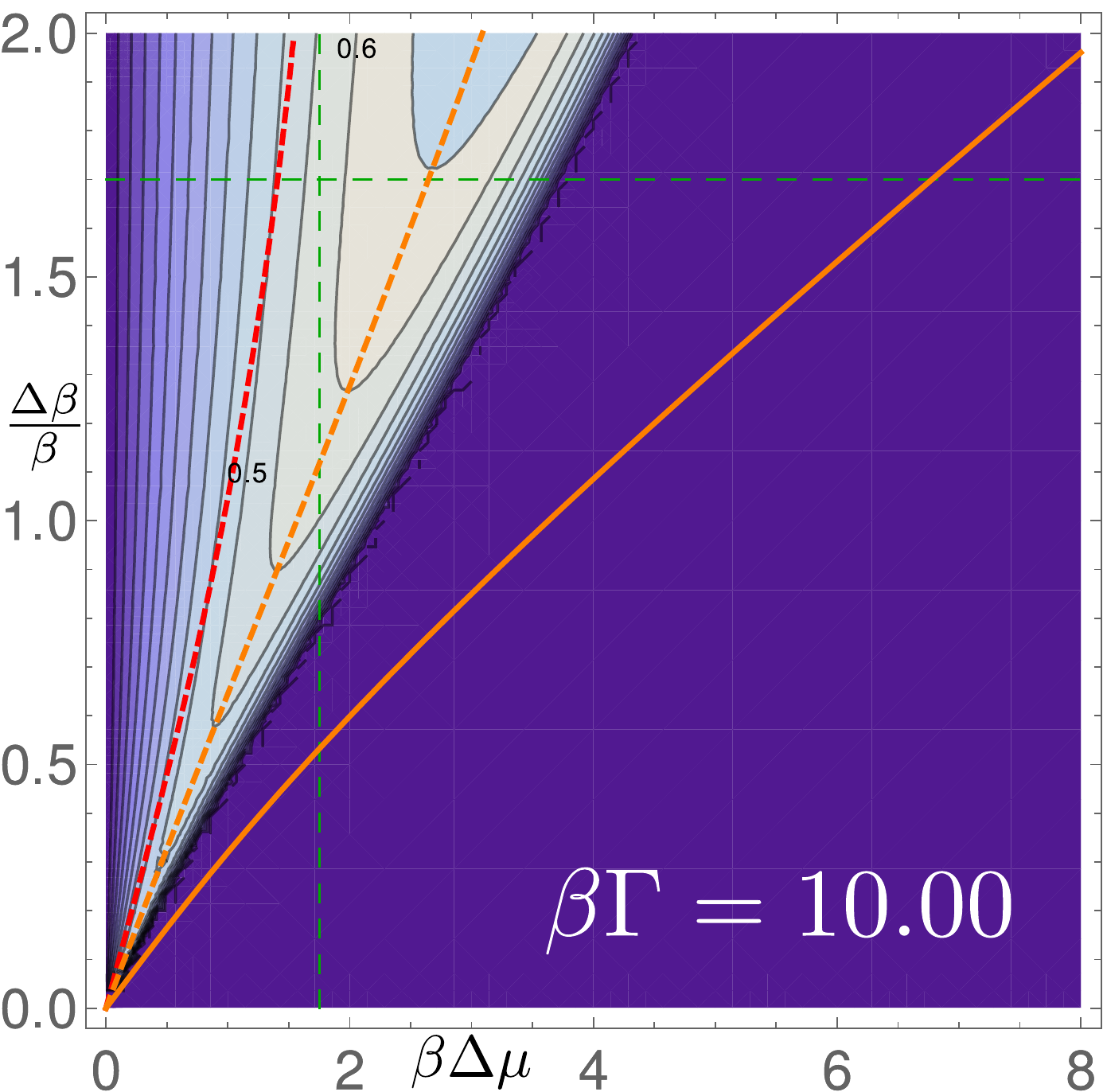} &
\includegraphics[width=0.3\textwidth]{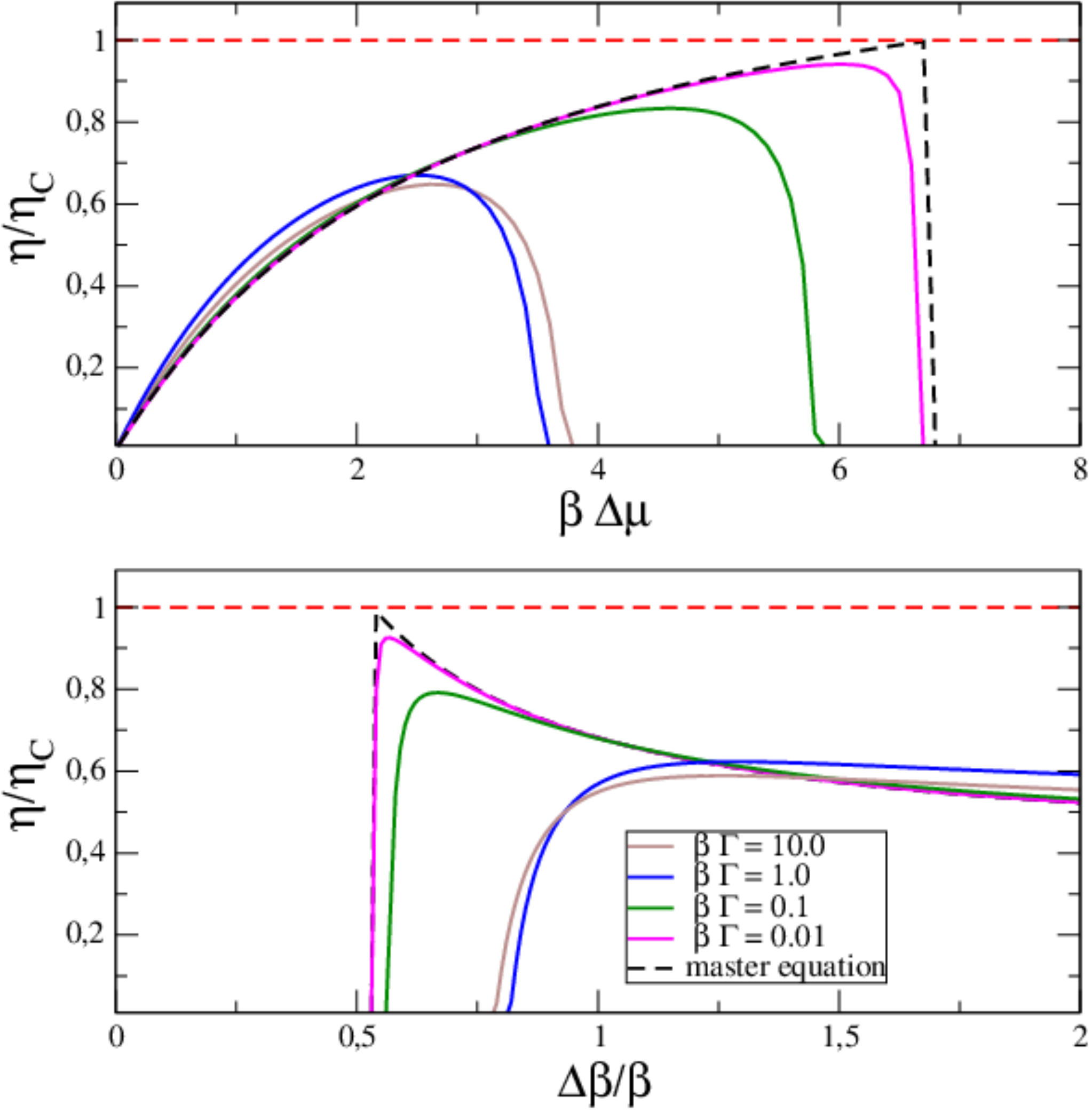}
\end{tabular}
\caption{\label{fig:figure}(Color Online)
Plot of the renormalized fermionic heat-to-power-conversion efficiency $\eta/\eta_{\rm C}$
versus dimensionless bias voltage and 
dimensionless temperature difference for different regimes.
In the weak coupling regime (${\beta\Gamma = 0.01}$, left panel) the efficiency reaches its maximum (dashed orange) versus the bias voltage near the
master equation prediction of zero power (solid orange).
The efficiency at maximum power (maximized versus bias voltage, dashed red) is far away from the maximum efficiency and thus 
significantly smaller than Carnot efficiency.
For strong coupling (${\beta\Gamma = 10.00}$, middle panel) efficiency at maximum power 
(dashed red) and maximum efficiency (dashed orange)
are much closer, and the master equation curve for zero power (solid orange) now yields a wrong 
bound for the region of finite efficiency.
The right panel shows for fixed temperature difference ${\Delta \beta/\beta = 1.7}$ (top, horizontal dashed line in contour plots) and for
fixed potential difference $\beta\Delta\mu=1.75$ (bottom, vertical dashed line in contour plots) that
for increasing coupling strength the maximum efficiency decreases.
Furthermore, it can be seen that also for small coupling strengths, the exact solution predicts that the maximum efficiency is reached
slightly before the power vanishes (magenta).
Contours denote efficiency steps of $0.05$.
Symmetric Lorentzian tunneling rates ${\Gamma_\text{L}(\omega) = \Gamma_\text{R}(\omega) = \Gamma \delta^2 /(\omega^2 + \delta^2)}$ 
with ${\delta = \epsilon}$ have been used throughout. 
Other parameters: ${\beta\epsilon=+2}$ and ${\mu=-\epsilon/2}$.
} 
\end{figure*} 
In Fig.~\ref{fig:figure} we plot the efficiency versus the temperature and potential gradients in the weak (left) and strong 
(middle) coupling regimes and also provide a direct comparison of the different coupling strengths (right).
We note that in contrast to the master equation efficiency -- cf. Eqns.~(\ref{eq:eff}) and~(\ref{eq:weak_coupling_current})  -- 
the true efficiency also varies with respect to temperature.
It is visible that the maximum efficiency versus bias voltage will decrease with increasing coupling strength.
Whereas for small coupling (left) it is close to the master equation results and thus close to Carnot efficiency, 
it decreases drastically for larger couplings (middle).
We attribute this to the loss of tight-coupling in the strong-coupling regime.
As a more positive feature of the strong-coupling regime -- in particular for practical applications -- however, we also observe that the
curves of efficiency at maximum power and maximum efficiency are closer together and that
maximizing the efficiency does not require to accept zero power.
Unfortunately, also in the strong-coupling regime -- compare Eqns.~(\ref{eq:strong_coupling_current}) -- these curves
will not coincide, such that one still has find a tradeoff between maximum power and maximum efficiency~\cite{whitney2014a}.
The plots for bosonic transport (not shown) are  very similar. 

\section{Summary}\label{SEC:discussion}

Using a simple equation of motion technique in combination with Laplace transform methods, 
we have calculated the exact steady-state energy and matter currents of simple two-terminal transport models
both for fermions and bosons, yielding a Landauer-type representation.
The approach can be expected significantly more complicated both from a technical and conceptual perspective 
when applied for finite times.
When combining the resulting expressions into heat currents, we showed explicitly that the 
steady-state entropy production rate is positive.
Using this second law inequality, we discussed the efficiency of converting a thermal gradient
into power.
For our transport models we noticed that for increased system-reservoir coupling strength, the
maximum efficiency decreases.
Importantly, we also note that the power at maximum efficiency is finite, 
rendering the strong-coupling regime a potential candidate for practical applications.

\acknowledgments
Financial support by the DFG (SCHA 1646/3-1, SFB 910, GRK 1558) and discussions with M. Esposito and S. Gurvitz
are gratefully acknowledged.

\bibliographystyle{eplbib}
\bibliography{references}

\end{document}